\documentclass[letterpaper, 10 pt, conference]{ieeeconf}
\IEEEoverridecommandlockouts
\usepackage{cite}
\newcommand{\req}[1]{\eqref{#1}}

\usepackage[linesnumbered, ruled]{algorithm2e}
\usepackage{bm}
\usepackage{color}
\usepackage{subfigure}
\usepackage{comment}
\usepackage{amsmath,amssymb,amsfonts}
\usepackage{algorithmic}
\usepackage{graphicx}
\usepackage{textcomp}
\usepackage{xcolor}
\usepackage{lscape}
\usepackage{multicol}
\newcommand{\qedwhite}{\hfill \ensuremath{\Box}}

\newtheorem{problem}{Problem}
\newtheorem{case}{Case}
\newtheorem{remark}{Remark}

\definecolor{airforceblue}{rgb}{0.36, 0.54, 0.66}

\newtheorem{Specification Set}{Candidate Specifications} 
\def\BibTeX{{\rm B\kern-.05em{\sc i\kern-.025em b}\kern-.08em
    T\kern-.1667em\lower.7ex\hbox{E}\kern-.125emX}}
\begin{document}
\setlength\textfloatsep{8.45pt}
\setlength\intextsep{8.45pt}
\title{Neural Controller Synthesis for Signal Temporal Logic Specifications Using Encoder-Decoder Structured Networks
}
\author{Wataru Hashimoto, Kazumune Hashimoto, Masako Kishida, and Shigemasa Takai
\thanks{Wataru Hashimoto, Kazumune Hashimoto, and Shigemasa Takai are with the Graduate School of Engineering, Osaka University, Suita, Japan (e-mail: hashimoto@is.eei.eng.osaka-u.ac.jp, hashimoto@eei.eng.osaka-u.ac.jp, takai@eei.eng.osaka-u.ac.jp). Masako Kishida is with the National Institute of Informatics, Tokyo, Japan (email: kishida@nii.ac.jp).
This work is supported by JST CREST JPMJCR201, Japan and by JSPS KAKENHI Grant 21K14184.
}}

\maketitle

\begin{abstract}
In this paper, we propose a control synthesis method for signal temporal logic (STL) specifications with neural networks (NNs). 
Most of the previous works consider training a controller for only a given STL specification. 
These approaches, however, require retraining the NN controller if a new specification arises and needs to be satisfied, which results in large consumption of memory and inefficient training. 
To tackle this problem, we propose to construct NN controllers by introducing \textit{encoder-decoder} structured NNs with an attention mechanism. The \textit{encoder} takes an STL formula as input and encodes it into an appropriate vector, and the \textit{decoder} outputs control signals that will meet the given specification. 
As the encoder, we consider three NN structures: \textit{sequential}, \textit{tree-structured}, and \textit{graph-structured} NNs. All the model parameters are trained in an end-to-end manner to maximize the expected robustness that is known to be a quantitative semantics of STL formulae. 
We compare the control performances attained by the above NN structures through a numerical experiment of the path planning problem, showing the efficacy of the proposed approach. 
\end{abstract}



\section{Introduction}\label{sec:intro}
Methods for addressing traditional control objectives such as stabilization and tracking are widely investigated in the automatic control field and have provided significant value for plenty of industrial applications. However, in robotic applications such as autonomous driving, more sophisticated control methods are required to achieve safe and highly automated control of the systems. For instance, cars on a road may be required to visit multiple places in a specified order while following complex traffic rules, which might not be handled solely by the classical control methods mentioned above. Such complex operations are often handled by hierarchically designing the different levels of controllers in an ad-hoc manner \cite{CPS}, while such approaches require the designer to face complex interactions among the controllers.   

One of the alternative approaches to dealing with complex task specifications is to write the specifications with temporal logic languages such as Linear Temporal Logic (LTL) \cite{LTL} or Signal Temporal Logic (STL) \cite{STL} and then incorporate them into the control design. 
LTL is one of the most popular temporal logic languages for control synthesis since the control problem with LTL constraints can readily be converted into an automaton for which there exist well-studied methods for synthesizing closed-loop controllers \cite{LTL1, LTL2}. However, these approaches are subject to scalability issues when they face high-dimensional systems. Moreover, the usage of the LTL language is limited to high-level planners since the LTL is defined over discrete states (atomic proposition) and cannot specify the continuous behavior of the system.

Recently, STL, which is the focus of this paper, has attracted much attention in control. Different from LTL, STL can specify the temporal properties of real-valued signals (e.g., state trajectories produced from dynamical systems) and allows us to formulate a variety of complex tasks including time constraints.
Notably, the STL is equipped with quantitative semantics called \textit{robustness} \cite{robustness} which indicates how much a system trajectory satisfies a given specification. 

In many previous works, the STL control synthesis problem is effectively formulated as the optimization problem by using the robustness function \cite{MILP1, MILP2, MILP3, MILP4,MILP5, smooth1, smooth2, smooth3}.
The authors of the works \cite{MILP1, MILP2, MILP3,MILP4,MILP5} formulated a control problem with STL constraints as a Mixed-Integer Linear Program (MILP) and implemented it in a receding horizon manner. One of the issues with these methods is that they do not scale well with the specification complexity since MILP is NP-hard. Moreover, these methods cannot deal with nonlinear dynamics since MILP requires all the constraints to be linear. To tackle such problems, a number of techniques for smoothing the robustness function are considered to utilize the gradient-based method \cite{smooth1, smooth2, smooth3}. In these methods, the resulting optimization problem becomes a sequential quadratic program (SQP), thus most of the above issues seen in \cite{MILP1} can be avoided or mitigated, although SQP may still be intractable for real-time implementation.

The development of the differentiable robustness functions and programming language toolbox for computing the robustness such as STLCG \cite{stlcg} accelerates the use of the neural networks (NNs) for STL control synthesis in recent years \cite{FNN, RNN1, RNN2, semi}. 
In these methods, the NNs for representing control policy are trained offline and then used in an online control execution, which enables much faster online computation than the methods directly solving the optimization problem at each time step. 
In the work \cite{FNN}, the control policy is represented by a feed-forward NNs (FNNs) and is trained to maximize the robustness via adversarial training.
Instead of FNNs, Recurrent Neural Networks (RNNs) are used in \cite{RNN1, RNN2} to explicitly take into account the history-dependent nature of satisfaction of the STL. 
The controller in \cite{RNN1} is trained by supervised imitation learning with a large training dataset consisting of the trajectories obtained by solving optimal control problems with STL constraints.
This dataset construction is time-consuming and may be infeasible depending on the control problem because of the non-convexity of the optimization problem.
To mitigate this issue, the authors of \cite{semi} introduced a semi-supervised training scheme, which incorporates the deviation from trajectories generated by human experts as well as STL robustness into the loss function. 
Deep Reinforcement Learning (DRL) and Learning from demonstrations (Lfd) based synthesis methods are also investigated in the previous works of literature \cite{RL1, RL2, RL3, RL4, demonstration1, demonstration2}. 

In general, the afore-cited previous works of NN controller synthesis aim at satisfying only a given STL specification. 
Therefore, if new specifications, which are not considered in the training procedure, are needed to be satisfied, the user has to retrain the NN controller for each of these tasks, which leads to large consumption of computational resources and memory (the number of the parameters to be learned could increase as the number of candidate specifications increases). 
To account for this problem, in this paper, we propose a novel learning scheme for neural controller synthesis of STL specifications. Specifically, we employ encoder-decoder structured NNs, in which the \textit{encoder} directly takes the STL formula as an input and generates a vector corresponding to the given specification and the \textit{decoder} takes the state of the system and the vector generated by the encoder as inputs and generates control signals as outputs. As the encoder, we utilize the \textit{sequential} \cite{LSTM,GRU}, \textit{tree-structured} \cite{tree}, or \textit{graph-structured} \cite{GNN1,GNN2} NNs to process a given STL specification. Since the sequential NNs cannot capture the logical structure of the STL specifications \cite{stlcg} and the model needs to read the operator's range of influence from the auxiliary variables such as bracket pairs, tree- or graph-structured NNs which can automatically extract such information may achieve better performance or more efficient training. The decoder is structured by sequential NNs, which can memorize sequential information through hidden states and deal with the history-dependent nature of the STL specification satisfaction, similar to the previous works regarding STL control synthesis \cite{RNN1,RNN2}. Then, the parameters of both encoder and decoder NNs are trained in an end-to-end manner by maximizing the expected robustness against the specifications for the training. 

\textbf{Contributions:}
The contributions of our work are as follows. 
First, we propose a method to synthesize NN controllers for STL specifications with encoder-decoder structured NNs aiming at generalizing the NN controller to different STL specifications.
As mentioned above, we consider three types of NN architectures (i.e., \textit{sequential}, \textit{tree-structured}, and \textit{graph-structured} NNs) with attention mechanisms and consider the corresponding training procedures. 
All the NN parameters are trained in an end-to-end manner to maximize the expected value of the robustness, which is known to be a quantitative semantics of STL formulae. 
Then, in the case study, we test the control performance of the resulting controller for a wide range of STL specifications and show the efficacy of the proposed method. Moreover, we compare the performances attained by all the NN structures of the encoder. 

\textbf{Related works:}
This study is built on the works regarding NN-based STL control synthesis methods \cite{FNN,RNN1} and considers generalizing these methods to multiple different STL specifications.
Some existing works have partially achieved such generalizations. In the works \cite{RNN1, RNN2, CBF}, the resulting NN controllers can deal with changing obstacles by utilizing the Control Barrier Function (CBF) in the training procedures. 
Another work \cite{semi} considers generalizing the controller by conditioning the control policy with an environment summary vector generated by Convolutional Neural Networks (CNNs) and enables the controller to deal with new environments without the need for re-synthesis of the NN controller.
However, the ability of these methods to deal with changes in task specifications is basically limited to obstacle avoidance or adaptation for new environments and these methods do not consider fully accounting for the changes in the STL specification itself. 
Inspired by the concept of word2vec \cite{skip-gram,word2vec}, our previous work \cite{STL2vec} proposed an approach to constructing "STL2vec" which converts STL specifications to latent representations that capture the similarities among them, and using them to construct a control policy. Although this method enables synthesis for multiple specifications and has the potential to significantly save memory consumption required for the training,
it has the drawback that the controller trained by \cite{STL2vec} cannot produce meaningful control signals when a specification not considered in the training is fed to the controller. 
The synthesis method proposed in this paper can overcome these limitations in the previous methods because of the NN architecture that can directly take a logical formula of STL as input.

This study is also related to symbolic logic embedding \cite{embedding1, embedding2, tree em, tree em2, embedding3, embedding4, embedding5,embedding6,embedding} that considers the methods for mapping logical formulae to real-valued vectors to incorporate high-level structured knowledge into NNs. 
In the works \cite{embedding1,embedding2}, logical formulae are directly taken as sequences, and LSTM with attention mechanism is used to process logical premises and hypotheses similar to the common procedure employed in natural language processing tasks. However, the complex and structured natures of logical formulae make this process challenging. To address this issue, graph-based methods \cite{embedding3, embedding4,  embedding5,embedding6,embedding} and tree-based methods \cite{tree em, tree em2} are developed to capture logical information. 
Referring to these previous results, we construct the encoder based on sequence, tree, and graph-structured NNs and compare the control performance attained by them.

Moreover, the generalization of the controller for unseen LTL tasks with deep reinforcement learning is considered in the previous works \cite{gene LTL1, gene LTL2, LTL2Action}. Different from these methods, in our work, the controller is constructed by an encoder-decoder structured NNs similar to the ones exploited for general sequence-to-sequence \cite{Seq2Seq,Attention} or graph-to-sequence \cite{Graph2Seq,Graph2Seq2,Graph2Seq3} tasks (e.g., machine translation, text generation) and can be efficiently trained in an end-to-end manner, thanks to the differentiable loss defined by the smooth robustness function mentioned above (since defining the differentiable loss for LTL tasks is basically challenging and incompatible with gradient-based methods, RL is suited to synthesize them while our method does not need to use RL if the system dynamics is known or can be predicted). 
In addition, as we will see in the case study in Section \ref{case study}, the resulting controller can flexibly handle a variety of specifications with time constraints, which cannot be handled by the LTL control synthesis methods.

\section{Preliminaries}\label{pre}
\subsection{System description and notations}\label{system}
We consider a nonlinear discrete-time dynamical system of the form: 
\begin{align}\label{dynamics}
    x_{t+1} = f(x_t, u_t),\ u_t\in \mathcal{U},
\end{align}
where $x_t\in \mathbb{R}^n$ is the system state at time $t\in \mathbb{Z}_{\geq 0}$, $u_t\in \mathcal{U}\subset \mathbb{R}^m$ is the control input at time $t$, and $f: \mathbb{R}^n \times \mathcal{U} \rightarrow \mathbb{R}^n$ is a function capturing the dynamics of the system. We assume that the initial state $x_0$ is randomly chosen from $\mathcal{X}_0\subset \mathbb{R}^n$ according to the probability distribution $p : \mathcal{X}_0 \rightarrow \mathbb{R}$, and $\mathcal{U}$ is defined by $\mathcal{U} = \{u \in \mathbb{R}^m : u_{\min}\leq u \leq u_{\max} \}$ for given $u_{\max}, u_{\min} \in \mathbb{R}^m$ (the inequalities are element-wise). 
Given $x_0 \in \mathcal{X}_0$ and a sequence of control inputs $u_0, \ldots, u_{T-1}$ with a horizon length $T$, we can generate a unique sequence of states according to the dynamics \req{dynamics}, which we call a \textit{trajectory}: $x_{0:T} = (x_0, x_1, \ldots, x_T)$. 
\color{black}
\subsection{Signal Temporal Logic}\label{STL}
In this subsection, we briefly summarize the basics of the Signal Temporal Logic (STL) \cite{STL}. STL is defined over signals (in this study, the signal is the state trajectory $x_{0:T}$ defined in Section \ref{system}). 
The syntax or grammar of the STL formula is recursively defined as follows:
\begin{align}\label{grammar}
    \phi ::= &\top \mid \mu \mid \neg \phi \mid \phi_1 \land \phi_2 \mid \phi_1 \lor \phi_2 \mid  \phi_1 \bm{U}_{I} \phi_2 
\end{align} 
where $\mu : \mathbb{R}^n \rightarrow \mathbb{B}$ is the predicate whose boolean truth value is determined by the sign of a function $h : \mathbb{R}^n \rightarrow \mathbb{R}$ defined over the system state (i.e., $\mu$ is true if $h(x_t)>0$, and false otherwise), $\phi$, $\phi_1$, and $\phi_2$ represent STL formulae, $\top$,  $\neg$, $\land$, and $\lor$ are Boolean \textit{true}, \textit{negation}, \textit{and}, and \textit{or} operators, respectively, and $\bm{U}_{I}$ is the temporal \textit{until} operator defined on a time interval $I =[a, b] = \{ t \in \mathbb{Z}_{\geq 0}: a\leq t \leq b\}$ ($a$, $b\in \mathbb{Z}_{\geq 0}$). An STL formula $\phi$ is generated by selecting an element from the list (\ref{grammar}) in a recursive manner.
Then, we define the \textit{Boolean} semantics of an STL formula $\phi$ with respect to the system trajectory starting from time $t$ (i.e., $x_{t:T}$) as follows:  
\begin{align}
    &x_{t:T} \models \mu \Leftrightarrow h (x_t)>0\notag \\
    &x_{t:T}\models \neg \mu \Leftrightarrow \neg (x_{t:T}\models \mu) \notag\\
    &x_{t:T}\models \phi_1 \land \phi_2 \Leftrightarrow x_{t:T}\models \phi_1 \land x_{t:T}\models \phi_2 \notag \\
    &x_{t:T}\models \phi_1 \lor \phi_2 \Leftrightarrow x_{t:T} \models \phi_1 \lor x_{t:T} \models \phi_2 \notag \\
     & x_{t:T}\models \phi_1 \bm{U}_{I}\phi_2 \Leftrightarrow \exists t_1 \in t+I\  \mathrm{s.t.}\  x_{t_1:T}\models \phi_2 \notag \\
     & \qquad \qquad \land \forall t_2 \in [t,t_1], x_{t_2:T}\models \phi_1\notag,
\end{align}
where $t+ I = \{t+k \in \mathbb{Z}_{\geq 0}: k \in I \}$. We note here that the trajectory length $T$ needs to be large enough to evaluate whether the specification $\phi$ is satisfied with the trajectory $x_{t:T}$ or not because of the definition above.
Stated in words, $\phi_1 \bm{U}_{I} \phi_2$ means that ``$\phi_2$ holds for the signal within a time interval $I$ and $\phi_1$ must always be true  against the signal prior to that". 
Other temporal operators $eventually$ and $always$ ($\bm{F}_{I}$ and $\bm{G}_{I}$) are defined based on the until operator as $\bm{F}_{I}\phi:= \top \bm{U}_{I}\phi$ and $\bm{G}_{I}\phi:=\neg \bm{F}_{I}\neg \phi$ respectively.  $\bm{F}_{I}\phi$ states that ``$\phi$ must hold at some time point within the interval $I$" while $\bm{G}_{I}\phi$ states that ``$\phi$ must hold for the signal within $I$". 

The notion of \textit{{robustness}} in STL provides \textit{{quantitative}} semantics, and it measures {how much} the trajectory satisfies the STL formula \cite{robustness}. The robustness is sound in the sense that positive robustness value implies satisfaction and negative robustness implies violation of the given STL formula. The robustness score of the STL formula $\phi$ over a trajectory $x_{t:T}$ is inductively defined as follows: 
\begin{align}
    & \rho^\mu(x_{t:T})&& = h (x_t) \notag \\
    & \rho^{\neg \mu}(x_{t:T}) &&= -h (x_t)\notag \\ 
    & \rho^{\phi_1 \land \phi_2}(x_{t:T}) &&= \min (\rho^{\phi_1} (x_{t:T}), \rho^{\phi_2} (x_{t:T}))\notag \\
    & \rho^{\phi_1 \lor \phi_2}(x_{t:T}) &&= \max (\rho^{\phi_1} (x_{t:T}), \rho^{\phi_2} (x_{t:T}))\notag \\
    & \rho^{\bm{F}_{I}\phi}(x_{t:T}) &&= \max_{t_1\in t+I}\rho^\phi(x_{t_1:T})\notag \\
    & \rho^{\bm{G}_{I}\phi}(x_{t:T}) &&= \min_{t_1\in t+I}\rho^\phi(x_{t_1:T})\notag \\
     & \rho^{\phi_1 \bm{U}_{I}\phi_2}(x_{t:T}) &&= \max_{t_1\in t+I}\Bigl(\min (\rho^{\phi_2}(x_{t_1:T}), \notag  \\
    & && \qquad \quad \min_{t_2 \in [t, t_1]}\rho^{\phi_1}(x_{t_2:T}))\Bigr).\notag
\end{align}
Same as the Boolean semantics, the trajectory length $T$ should be large enough to determine the robustness score. 
\begin{remark}\label{remark:smooth}
Due to the definition above, the robustness function is generally non-differentiable since it can be nested with non-differentiable max/min functions. Thus, we cannot directly use gradient-based methods with the original robustness above.
To account for this problem, we adopt a \textit{smooth approximation} of the min/max operators by the log-sum-exp as follows: $\max (a_1,\dots,a_m) \approx \frac{1}{\beta}\mathrm{ln} \sum_{i=1}^m \exp (\beta \alpha_i)$ and $\min (a_1,\dots,a_m) \approx \frac{1}{\beta}\mathrm{ln} \sum_{i=1}^m \exp (-\beta \alpha_i)$,
where $\beta >0$ is the scaling parameter. When $\beta\rightarrow \infty$, the approximation approaches the true robustness value \cite{approx4}. 
\qedwhite 
\end{remark}
\color{black}

\section{Problem Statement}\label{sec:problemstatement}
We assume that the function $f$ of the system (\ref{dynamics}), the probability distribution of initial states $p$, the horizon length $T$, the set of STL specifications $\Phi$, and the probability distribution from which an STL specification is sampled (i.e., $\phi \sim p_{\phi}$) are given (concrete examples of $p_\phi$ are discussed in Section~\ref{case study}). 
Let a control policy be given by  $\pi(\cdot,\cdot):\mathbb{R}^{n(t+1)} \times \Phi \rightarrow \mathbb{R}^{m}$, whose inputs are any $\phi\in \Phi$ and a sequence of the states including the current and the past time steps $x_{0:t} = (x_0, \ldots, x_t)$, and output is a control signal to be applied for the current time, i.e., $u_t = \pi (x_{0:t}, \phi)$. 
Before control execution, a pair of the initial state $x_0$ and the specification $\phi \in \Phi$ is sampled, and then the trajectory is generated according to the policy $\pi(\cdot,\phi)$, 
i.e., $x_0\sim p$, $\phi \sim p_\phi$, and $x_{t+1} = f(x_{t},u_t)$ with
$u_t = \pi(x_{0:t},\phi)$, $t=0, 1, \ldots $.
Note that the control policy above is a function of current and past system states $x_{0:t}$ (instead of $x_t$) due to the history-dependent property of the STL specification satisfaction (see e.g., \cite{RNN1}). 

Our goal is to synthesize a control policy $\pi(\cdot,\cdot): \mathbb{R}^{n(t+1)}\times \Phi \rightarrow \mathbb{R}^{m}$, such that the following expected robustness is maximized: 
\begin{align}
 \mathbb{E}_{x_0 \sim p, \phi \sim p_\phi} \left [ \rho^{\phi} \left(x_{0:T}^{\pi_\phi}\right)\right], \label{optimization2}
\end{align}
where $x_{0:T}^{\pi_\phi} = \left (x_0^{\pi_\phi}, \ldots, x_T^{\pi_\phi}\right )$ with $x_0^{\pi_\phi} = x_0$ is the state trajectory obtained by applying the control policy $\pi (\cdot, \phi)$. 
In this paper, we aim at synthesizing this control policy based on a neural network (NN). To achieve this, we need to consider a concrete NN architecture 
that can directly take any STL formula $\phi \in \Phi$ and a trajectory $x_{0:t}$ as the inputs and a control signal $u_t$ as the output. Moreover, we need to consider a concrete training procedure of the entire NN parameters, such that the expected robustness is maximized according to \req{optimization2}. 
Our solution approach including the selections of the NN architecture and the corresponding training schemes will be discussed in the following section. 

\section{Proposed Method}\label{sec:proposed}
\begin{figure*}[tb]
 \begin{center}
  \includegraphics[width=1\hsize]{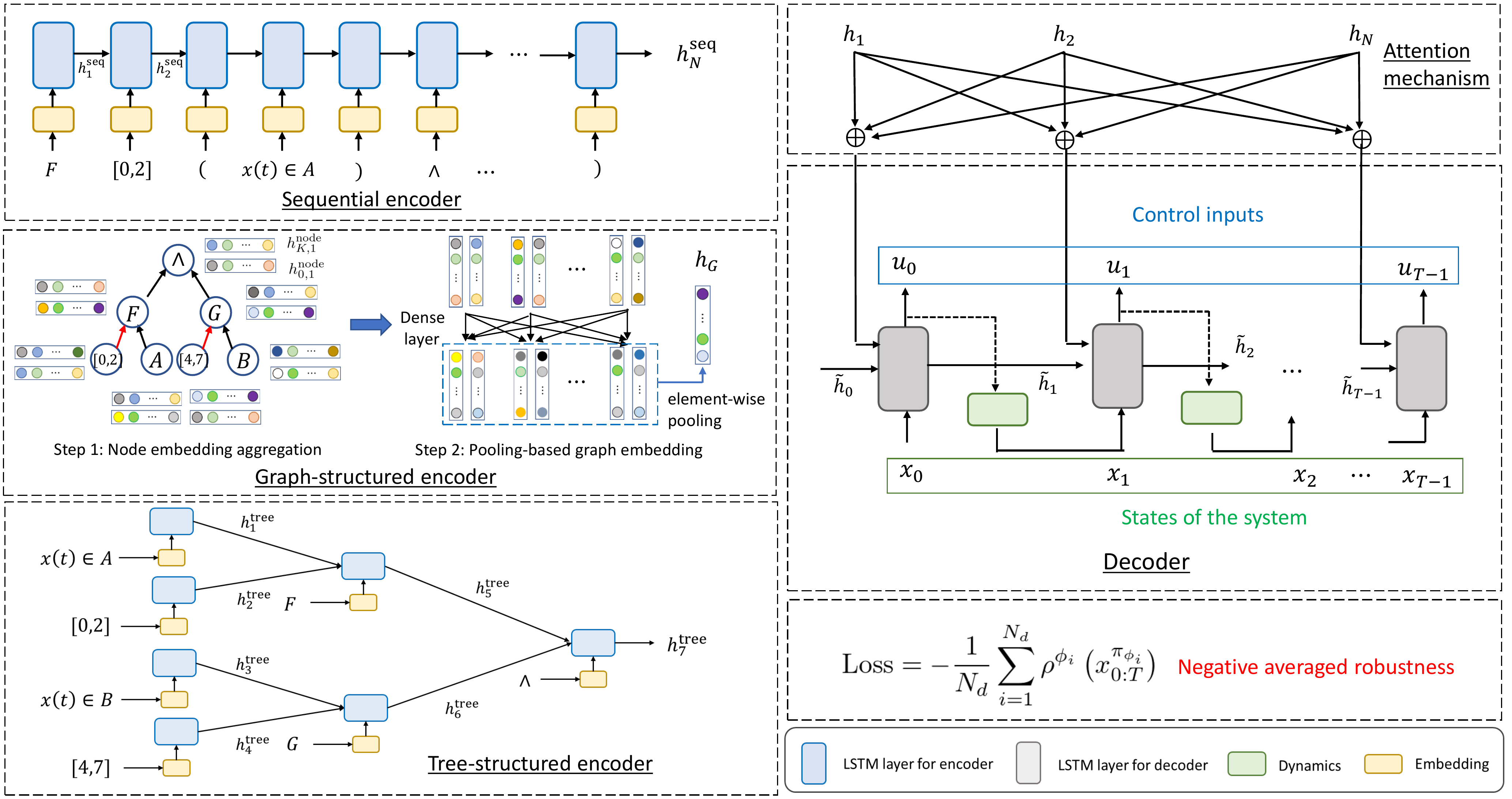}
 \end{center}
 \caption{Proposed encoder-decoder model structure for STL control synthesis: The vectors that represent every element within a given STL formula are fed to the encoder that is constructed by sequential, tree-structured, or graph-structured NNs and generates specification summarized vector and hidden vectors. Then, the vectors are passed to the decoder. The decoder generates control signals based on the vectors received from the encoder and the state feedback from the system. The initial hidden state of the decoder is defined by the specifications summarized vector from the encoder. In addition, the attention mechanism is implemented by using the hidden vectors from the encoder to improve the training performance. In this figure, the example with the specification $F_{[0,2]}(A)\land G_{[4,7]}(B)$ (where $A$ and $B$ are minimum units of STL specifications such as predicate) is shown.}
 \label{fig:proposed}
\end{figure*}

To achieve the goal discussed in Section \ref{sec:problemstatement}, we construct an encoder-decoder structured NN controller that directly takes any STL formula $\phi \in \Phi$ as input, encodes it, and generates appropriate control signals that satisfy $\phi$. The overview of the proposed encoder-decoder NN architecture is summarized in Fig. \ref{fig:proposed}.
The proposed NN controller is trained in an end-to-end manner, aiming at minimizing a certain loss (defined later in Section IV-B). 
In the following, we explain the concrete model architectures employed in this study and the detailed training procedure of them in Sections IV-A and IV-B respectively. 
\subsection{Controller architecture}\label{model}
In this subsection, we discuss the concrete architecture of the proposed NN controller. The encoder and decoder NNs are explained in Section \ref{encoder} and \ref{decoder}, respectively.
\subsubsection{Encoder}\label{encoder}
The role of the encoder is to read and encode an STL formula into a continuous vector representation so that it can be passed to the decoder.
In this study, we consider three types of model structures for the encoder: \textit{sequential model}, \textit{graph-structured model}, and \textit{tree-structured model}. In what follows, we explain each of them in detail.

\textbf{Sequential encoder:}
Referring to commonly used architectures in natural language processing \cite{Seq2Seq, Attention}, we first consider constructing an encoder with sequential NNs. In this architecture, a sequence of vectors $s_{1:N}=(s_0,s_1,\ldots,s_{N-1})$ representing the given STL formula is fed to a sequence model such as Long Short Term Memory (LSTM), where vectors $s_i\ (i=0,2,\ldots, N-1)$ are obtained by simply encoding each component within the STL formula (i.e., predicates, logical or temporal operators, time bounds, and brackets that indicate operator's range of influence) into a prespecified vector (See Section~V for a more concrete example) and $N$ is a number of elements within a specification.
Then, the model sequentially processes the input vectors $s_i$ by the following formulation:  
\begin{align}
    h_{i+1}^{\mathrm{seq}} = g_{W_1^{\mathrm{seq}}}(h^{\mathrm{seq}}_{i},s_i),\ i = 0,1,\ldots, N-1,
\end{align}
where $h^{\mathrm{seq}}_i$, $i = 0,1,\ldots, N-1$ represent hidden states that memorize sequential information and $g_{W_1^{\mathrm{seq}}}$ is a non-linear function parameterized by $W_1^{\mathrm{seq}}$.
After processing all the input vectors, the last hidden vector $h_N^{\mathrm{seq}}$ is handed over to the decoder as a specification summarized vector.
Furthermore, all of the hidden states $h_{\mathrm{all}}^{\mathrm{seq}}=(h_1^{\mathrm{seq}},h_2^{\mathrm{seq}},\ldots, h_{N}^\mathrm{seq})$ are also sent to the decoder to implement so-called attention mechanism \cite{Attention} which is explained in the next subsection.

\textbf{Graph-structured encoder:}
The aforementioned sequential encoder has a potential drawback that it cannot capture the logical structure of the STL specifications without memorizing various auxiliary information (e.g., bracket pairs). Therefore, as given STL specifications become more nested or long, accurately extracting the relative relationships among specifications in terms of control becomes a more complex endeavor. 
Since the recursive definition of the STL semantics can be represented by parse trees whose each node represents each element in the given STL specification such as logical/temporal operators and predicates \cite{stlcg}, utilizing the Graph Neural Networks (GNNs) \cite{GNN1,GNN2} is one of the promising ways to deal with this problem. Indeed, the usefulness of the GNNs for logical formula embedding is shown in the previous works regarding compositional embedding \cite{embedding3, embedding4,embedding5,embedding6, embedding} and the generalization for LTL tasks \cite{LTL2Action}. 

In this study, we consider an architecture similar to the encoder part of the Graph2Seq model \cite{Graph2Seq} which is proposed mainly for natural language processing tasks. First, we convert the given STL specification into the corresponding graph representation $\mathcal{G}=\{ \mathcal{V}, \mathcal{E}\}$, where $\mathcal{V}$ and $\mathcal{E}$ are the set of nodes and edges within the graph, respectively. We denote the set of the incoming neighbor nodes of a node $v\in \mathcal{V}$ as $\mathcal{N}_{\mathrm{in}}(v)$. Each node $v\in \mathcal{V}$ represents an operator, time bounds, or predicates within the given formula. As shown in the example in Figure \ref{fig:proposed}, the nodes that represent time bounds and predicates are the leaf nodes that have the outgoing edge directed to the corresponding temporal operator and outer operator, respectively. The nodes that represent logical or temporal operators are the root nodes or the intermediate nodes that have the incoming edge from the subformulae and/or node of time bounds. Then, the graph encoder considered in this study first generates node embeddings that are the vectors assigned to all of the nodes and then constructs graph embedding that summarizes the given STL specification based on all of the learned node embeddings. 
The detailed generation processes of the node embeddings and graph embedding are the followings. First, we assume that all of the nodes $v_i\in \mathcal{V}$ with $i=1,2,\ldots, N_g$ have their initial feature vectors (initial embeddings) $h^{\mathrm{node}}_{0,i}$, which are defined by mapping the mutually identifiable vectors that represent each component of the STL specification similar to the sequential model case to user-specified dimensional vector space by dense layer with parameter $W_1^G$. Then, each node aggregates its own features $h_{t,i}^{\mathrm{node}}$ and the incoming neighbors' features $h_{t,j}^{\mathrm{node}}$ with $j=\{ j\mid v_j \in \mathcal{N}_{\mathrm{in}}(v_i) \}$ and updates its embedding. This aggregation step is implemented by the following:
\begin{align}\label{agg}
   h_{t,i}^{\mathrm{node}} =  \sigma \left(W_2^G \bar{h}_{t-1,i}^{\mathrm{node}} +  \sum_{r\in \{0,1,2,3\}} \sum_{j\in \mathcal{N}_{\mathrm{in}}(v_i)}W_{r,3}^G \bar{h}_{t-1,j}^{\mathrm{node}}\right),
\end{align}
where $\sigma$ is a nonlinear activation function, $\bar{h}_{t,i}^{\mathrm{node}}$ is the concatenation of the initial embedding ${h}_{0,i}^{\mathrm{node}}$ and the embedding at $t$-th aggregation step ${h}_{t,i}^{\mathrm{node}}$ (we have observed that the performance improves by using this concatenation instead of directly using ${h}_{t,i}^{\mathrm{node}}$ in the aggregation step), $W_2^G$ and $W_{r,3}^G$ ($r=\{0,1,2,3\}$) are the weight matrices to be trained. Since the relationship between a time-bound node and corresponding temporal operator node intuitively has different nature from that of the others, the weight matrices are separately trained for this relationship (i.e., for the edges between a time-bound node and temporal operator node, $r=1$ is assigned and for the other edges, $r=0$). We found that this improves the resulting control performance. Moreover, since we need to distinguish the left and right-hand side sub-formulae of the \textit{until} operator, the weight matrices for these relations are also separately trained (i.e., $r=2$ for the relation between the left sub-formula fed to \textit{until} operator and $r=3$ for the relation between the right sub-formula). 
After implementing this aggregation process $K$ time steps, we construct graph embedding based on the resulting node embeddings to obtain a more compact representation of the graph. Although we can use a variety of the down-sampling strategies as mentioned in Section IV-C of \cite{GNN2}, we here specifically use max/mean/sum pooling-based strategy that all of the node embeddings are fed to a fully connected neural network and applied max/mean/sum operation element-wise. 
\begin{align}
    h_{G} = max / mean / sum \left(\gamma_{W_4^G}(\bar{h}_{T,1}^{\mathrm{node}}),\ldots, \gamma_{W_4^G}(\bar{h}_{T,N_g}^{\mathrm{node}})\right),
\end{align}
where $\gamma_{W_4^G}$ is a dense layer parameterized by $W_4^G$.
Finally, the graph embedding $h_G$ and all the node embeddings at the last aggregation step $h_{\mathrm{all}}^{\mathrm{node}}=(h_{T,1}^{\mathrm{node}},h_{T,2}^{\mathrm{node}},\ldots, h_{T,N_g}^{\mathrm{node}})$ are sent to the decoder as the specification summarized vector and vectors used for attention mechanism respectively.

\textbf{Tree-structured encoder:}
We also consider the tree-structured encoder. Different from the synchronized node aggregation step in GNNs (\ref{agg}), the update rule of the tree-structured encoder introduced here processes the hidden vectors in a bottom-up manner. Since the STL formula has a bottom-up tree structure, this feature may lead to better performance compared to the GNN-based encoder. 
We specifically employ the model structure based on Tree LSTM \cite{tree}. The update equation of the hidden state can be conceptually written as the following:
\begin{align}
    h_{i+1}^{\mathrm{tree}} = \bar{g}_{W_1^{\mathrm{tree}}}\left(\{h^{\mathrm{tree}}_{j}\}_{j\in C(i)},s_i\right),\ i = 0,1,\ldots, N-1,
\end{align}
where $C(i)$ is the set of all the child nodes of node $i$ and $\bar{g}_{W_1^{\mathrm{tree}}}$ is a non-linear function parameterized by $W_1^{\mathrm{tree}}$. Then, the last hidden state $h_N^{\mathrm{tree}}$ and set of all the hidden vectors $h_{\mathrm{all}}^{\mathrm{tree}}=(h_1^{\mathrm{tree}},h_2^{\mathrm{tree}},\ldots, h_N^\mathrm{tree})$ are sent to the decoder as the specification summarized vector and vectors for attention mechanism respectively.
The detailed update equations are shown in Appendix \ref{ape:treelstm}.
\subsubsection{Decoder}\label{decoder}
The role of the decoder is to generate a control input $u_t$ based on the current and past system states $x_{0:t}$ and the specification summarized vector received from the encoder (i.e., $h_N^{\mathrm{seq}}$ for the sequence encoder case or $h_G$ for the graph-structured encoder case, and $h_N^{\mathrm{tree}}$ for the tree-structured encoder case). To account for the long-term dependency of the control policy (i.e., control policy depends on the past system states), we again employ sequential NNs such as LSTM which can convey information regarding the past inputs (past system states) through the hidden states. We denote the hidden state of the decoder at time $t$ as $\tilde{h}_t$ to distinguish it from that of the encoder. The initial hidden state of the decoder $\tilde{h}_0$ is defined by the task embedding received from the encoder (i.e., $\tilde{h}_0=h_N^{\mathrm{seq}}$, $\tilde{h}_0=h_G$, or $\tilde{h}_0=h_N^{\mathrm{tree}}$). Moreover, we also employ the attention mechanism proposed in \cite{Attention} which uses all the hidden states $h_{\mathrm{all}}^{{\mathrm{seq}}}$ (sequence encoder case), $h_{\mathrm{all}}^{{\mathrm{tree}}}$ (tree encoder case), or node embeddings $h_{\mathrm{all}}^{{\mathrm{node}}}$ (graph encoder case) generated by the encoder to add more flexibility to the decoding process. The attention mechanism is considered to be effective in our work since intuitively, the desired control input at each time step will strongly depend on some specific portions of the given STL specification (e.g., for the specification $\phi= F_{[0,10]}(\cdot)\land F_{[10,20]}(\cdot)$, the controller should pay more attention to the portion $F_{[10,20]}(\cdot)$ than $F_{[0,10]}(\cdot)$ after $t=10$).
The update rule of the hidden state $h_t$ and calculation of the control input at time $t$ with the attention mechanism are as follows: 
\begin{align}
    \tilde{h}_{t+1} = \tilde{g}_{\tilde{W}_1} (\tilde{h}_{t}, x_t),\ u_t= l_{\tilde{W}_2} (\tilde{h}_{t}, z_t), \label{hteq} 
\end{align}
where $\tilde{g}_{\tilde{W}_1}$ represents non-linear function with parameter $\tilde{W}_1$, $l_{\tilde{W}_2}$ is an output layer parameterized by a weight $\tilde{W}_2$, and $z_t$ denotes so cold \textit{context vector} defined by the weighted sum of the vectors $(h_1,h_2,\ldots,h_N)$ within $h_{\mathrm{all}}^{\mathrm{seq}}$ or $h_{\mathrm{all}}^{\mathrm{node}}$ received from the encoder as follows:
\begin{align}
    z_t = \sum_{j} \alpha_{t,j}h_j,\ \  \alpha_{t,j} = \frac{\exp(\beta_{t,j})}{\sum_{j}\exp(\beta_{t,j})},
\end{align}
where $\alpha_{t,j}$ represents the attention weight that determines which part of encoder outputs should be referred to generate the control signal $u_t$. 
$\alpha_{t,j}$ is defined by the alignment term $\beta_{t,j}$ derived from the hidden states $h_j$ and $\tilde{h}_t$ as the following:
\begin{align}
    \beta_{t,j} = \tilde{w}^\top \tanh (\tilde{W}_3\tilde{h}_{t-1}+\tilde{W_4}h_j), 
\end{align}
where $\tilde{W}_3$, $\tilde{W}_4$, and $\tilde{w}$ are the hyperparameters to be learned. Note that hyperbolic tangent is applied element-wise.

Furthermore, same as the work \cite{FNN}, we employ hyperbolic tangent as the activation function of the output layer to restrict the produced control input $u_t$ within the lower bound $u_{\mathrm{min}}$ and the upper bound $u_{\mathrm{max}}$. The concrete operation of the output layer is as follows: 
\begin{align}
    u_{t} = u_{\mathrm{min}} + \frac{u_{\mathrm{max}}-u_{\mathrm{min}}}{2}\odot\left(\mathrm{tanh}\left(\tilde{W}_2[\tilde{h}_t^\top z_t^\top]^\top \right)+\mathbb{I}\right), \label{uteq}
\end{align}
where $\odot$ denotes element-wise multiplication and $\mathbb{I}$ is the vector whose elements are all 1.
Using \req{uteq}, we can generate control inputs satisfying $u_t \in \mathcal{U}$. 

\subsection{Training model parameters}
Given the encoder-decoder model structure discussed in Section \ref{model} the remaining question to achieve the goal discussed in Section \ref{sec:problemstatement} is how to train a set of all the NN parameters $W$ (i.e., a set of parameters for the encoder $W_1^{\mathrm{seq}}$ (sequential encoder case), $W_1^G,\ldots,W_4^G$ (graph encoder case), $W_1^{\mathrm{tree}}$ (sequence encoder case) and decoder $\tilde{W}_1,\ldots,\tilde{W}_3,\tilde{w}$). The whole training procedure is summarized in Algorithm \ref{alg}. 
Based on the discussion in Section \ref{sec:problemstatement}, we here consider finding a set of model parameters $W$ that solve the following maximization problem. 
\begin{align}
\mathop{\mathrm {maximize}}_{{W}} \ &  \mathbb{E}_{x_0\sim p, \phi \sim p_\phi} \left [ \rho^{\phi} \left(x_{0:T}^{\pi_\phi}\right)\right]. \label{optimization2}
\end{align}
Since the expectation in (\ref{optimization2}) cannot directly be evaluated, we approximately evaluate it with finite samples and update all the parameters end-to-end through back-propagation as we will see in the followings. Although the specifications used for training are limited to the distribution $p_\phi$, the trained model potentially can deal with the specifications out of the distribution because of the model structures that can take any specifications as input, which is discussed later in Section \ref{case study}. 
Moreover, since the NN controller might not guarantee the satisfaction of a newly given specification in the online execution phase, we additionally update the decoder parameters if a newly given specification cannot be satisfied with the current NN controller. 
The concrete procedure for the parameter update and adaptation for the new specification is summarized in the following subsections.

\subsubsection{Parameter update}
The entire model parameters are updated to solve the problem (\ref{optimization2}).
In each parameter update step, the dataset is constructed by the pairs of the initial state and STL specification that are randomly sampled from the distribution $p \times p_\phi$ as $\mathcal{D}=\{ (x_{0,i},\phi_i) \}_{i=1}^{N_d}$, where $N_d$ is the number of pairs used in each parameter update iteration.
Then, the NN parameters are updated using all the pairs $(x_{0,i}, \phi_i)$ within $\mathcal{D}$, via the following forward and backward computation. 
In the forward computation, the robustness values corresponding to every pair of initial state and specification $(x_{0,i}, \phi_i)$ in $\mathcal{D}$ (i.e., the robustness of $\phi_i$ for the state trajectory generated by alternately applying the control policy $\pi(\cdot,\phi_i; W)$ and system dynamics model $f$ from the initial state $x_{0,i}$) are computed. 
Then, we compute the following negative averaged robustness as a loss:
\begin{align}\label{ave robustness}
    \mathrm{Loss}=
    -\frac{1}{{N_d}} \sum_{i=1}^{{N_d}}  \rho^{\phi_{i}} \left(x_{0:T}^{\pi_{\phi_i}}\right).
\end{align}
When we calculate the robustness, we use STLCG toolbox \cite{stlcg} that uses computation graphs to calculate the robustness and can be well integrated with the existing auto-differentiation tools. 
After the forward computation, gradients $\Delta W$ of the averaged robustness (\ref{ave robustness}) with respect to all the parameters $W$ within the whole NN model (encoder and decoder) are computed by applying Back Propagation Through Time (BPTT). We can easily implement this procedure by using auto-differentiation tools designed for the NNs such as PyTorch. Then, finally, all the parameters are updated based on the obtained gradients using the existing optimizer such as adam \cite{adam}.
\begin{remark}\label{remark:smooth}
Since the smooth robustness function is non-convex, updating the parameters $W$ using the gradients of the loss (\ref{ave robustness}) may lead to a sub-optimal solution that does not satisfy the given specification, which is a common problem in STL control synthesis literature. To mitigate this problem, the scaling parameter $\beta$ mentioned in Remark 1 should be carefully chosen. More radical solutions to this problem would be considered in future work. 
\qedwhite 
\end{remark}

\subsubsection{Adaptation for newly given specifications}
After training the model parameters, we apply the learned control policy for a newly given specification $\phi$. 
Since the control performance obtained from the trained NN controller is not guaranteed to be optimal, we consider additionally updating the parameters of the decoder NN $\tilde{W}_1,\ldots, \tilde{W}_3,\tilde{w}$ based on the gradients of the negative robustness for the given specification $\phi$ using adam optimizer until the robustness for the given specification $\phi$ reaches the user-specified value $c>0$. 
If the training of the controller has been successfully done, the number of gradient steps required for this adaptation is expected to be much smaller than training the controller for the specification $\phi$ from scratch.

\begin{algorithm}[t]\label{alg}
{\small
\SetKwInOut{Input}{Input}
\SetKwInOut{Output}{Output}
\Input{$p$ (distribution of initial state); $p_\phi$ (distribution of STL training specification); $f$ (system dynamics model); $T$ (horizon length); encoder-decoder model structure; $N_{\mathrm{ite}}$ (maximum number of training iteration); $N_{d}$ (number of specifications considered in each iteration); $c$ (user-specified minimum required robustness value)}
\Output{$W$ (entire model parameters)} 
\textbf{[Parameter training]}\\
\For{$\ell = 1 : N_{\mathrm{ite}}$ }{

Sample $N_{d}$ pairs of initial state and specification $\mathcal{D}=\{ (x_{0,i}, \phi_i)\}_{i=1}^{N_{d}}$ from the distribution $p\times p_\phi$;\\
Compute the loss (\ref{ave robustness});\\
Implement back-propagation and obtain the gradient $\Delta W$ of the loss (\ref{ave robustness}) with respect to the model parameters $W$;\\
Update the parameters $W$ with $\Delta W$ using the Adam optimizer;\\
$\ell \leftarrow \ell+1$;\\}

\textbf{[Adaptation for given specification]}\\
$\phi$: newly given specification from the user;\\
\While{$\rho^{\phi} \left(x_{0:T}^{\pi_\phi}\right)<c$}{Obtain the gradients of the loss $-\rho^{\phi} \left(x_{0:T}^{\pi_\phi}\right)$ with respect to the decoder model parameters $\tilde{W}_1,\ldots, \tilde{W}_3,\tilde{w}$ thorough back-propagation;\\
Update the decoder parameters using the adam optimizer;}




    \caption{Model parameter training} 
    }
\end{algorithm}
\section{Case Study}\label{case study}

\begin{figure*}[htbp]
  \begin{minipage}{0.49\hsize}
    \begin{center}
      \includegraphics[width=1\hsize]{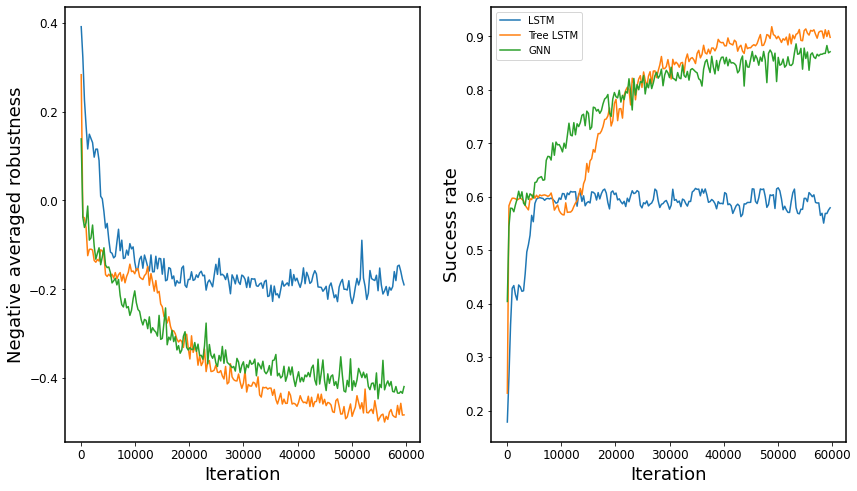}
    \end{center}
      \caption{Negative averaged robustness score and success rate (without attention mechanism)}\label{result:robustness1}
  \end{minipage}   
  \hspace{0.03\columnwidth}
  \begin{minipage}{0.49\hsize}
    \begin{center}
      \includegraphics[width=1\hsize]{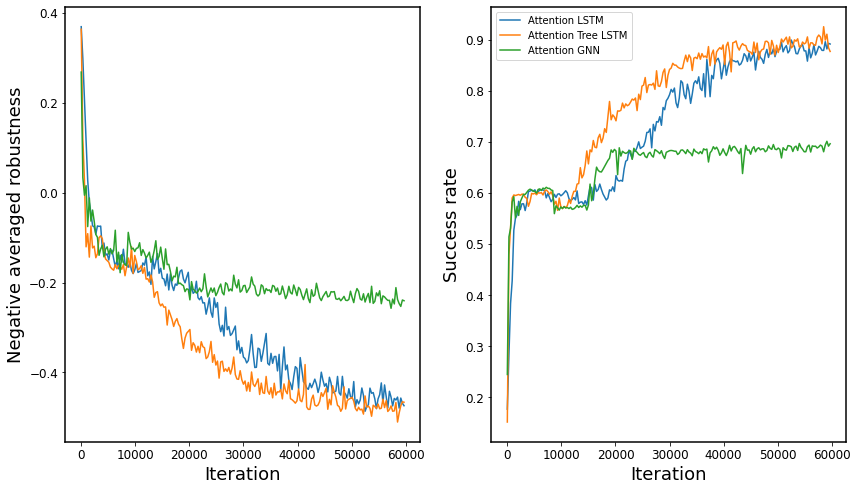}
    \end{center}
      \caption{Negative averaged robustness score and success rate (with attention mechanism)}\label{result:robustness2}
  \end{minipage}
\end{figure*}
\begin{figure*}[tb]
 \begin{center}
  \includegraphics[width=1\hsize]{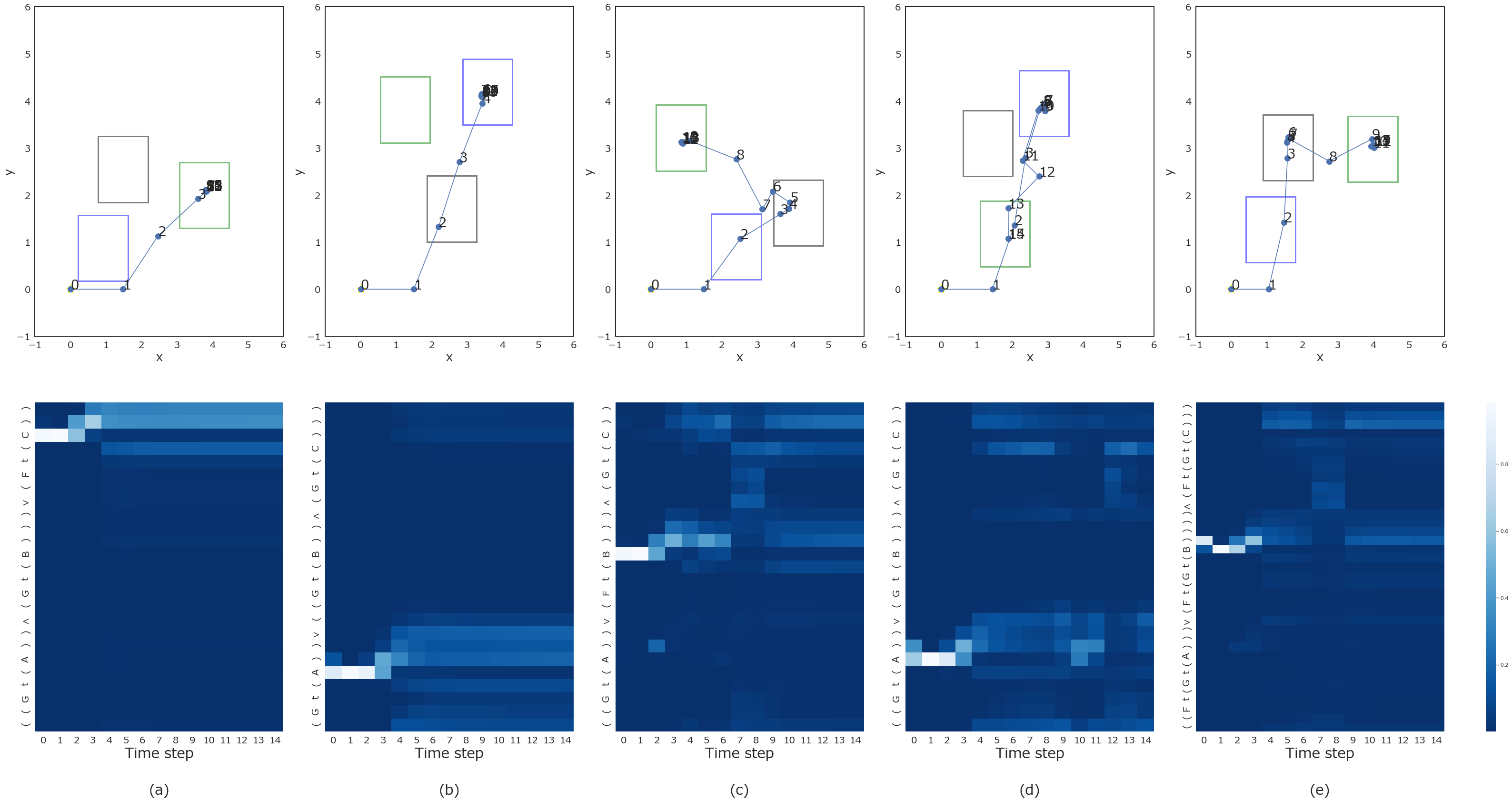}
 \end{center}
  \caption{A few examples of the trajectories generated by applying trained control policy to the system for some test specifications (above) and visualization of the attention weights for the sequential encoder (below). Test specifications are (a)$((\bm{G}_{[13,14]}(x_t \in \mathrm{A}))\land (\bm{G}_{[5,11]}(x_t \in\mathrm{B}))) \lor (\bm{F}_{[7,8]}(x_t \in \mathrm{C}))$ 
  (b)$(\bm{G}_{[4,9]}(x_t \in \mathrm{A}))\lor ((\bm{G}_{[3,8]}(x_t \in \mathrm{B})) \land (\bm{G}_{[11,13]}(x_t \in \mathrm{C})))$
  (c)$((\bm{G}_{[2,4]}(x_t \in \mathrm{A}))\lor (\bm{F}_{[4,12]}(x_t \in \mathrm{B}))) \land (\bm{G}_{[10,15]}(x_t \in\mathrm{C}))$
  (d)$((\bm{G}_{[8,9]}(x_t \in \mathrm{A}))\land (\bm{G}_{[2,10]}(x_t \in\mathrm{B}))) \lor (\bm{G}_{[14,15]}(x_t \in\mathrm{C}))$ (e)$\bm{F}_{[0,16]}\bm{G}_{[0,2]}(x_t \in \mathrm{A}) \land \bm{F}_{[0,16]}\bm{G}_{[0,2]}(x_t \in\mathrm{B}) \land \bm{F}_{[0,16]}\bm{G}_{[0,6]}(x_t \in \mathrm{C})$, 
  (d)$((\bm{F}_{[0,12]}\bm{G}_{[0,3]}(x_t \in \mathrm{A}))\lor (\bm{F}_{[0,13]}\bm{G}_{[0,2]}(x_t \in \mathrm{B}))) \land (\bm{F}_{[0,12]}\bm{G}_{[0,3]}(x_t \in \mathrm{C}))$,
  where A, B, and C represent blue, black, and green regions in figures respectively. The numbers indicated in the above figures represent time steps.
  In the visualizations of the attention weights (below figures), the vertical axis represents the given STL specification (sequence of the operators, time bounds, brackets, and predicates) and the horizontal axis represents discrete time steps. The values (color of the heat map) are attention weights in each time step for the encoder's hidden states corresponding to each element of the specification indicated in the vertical axis.
  Moreover, ($x_t\in A$), ($x_t\in B$), and ($x_t\in C$) are abbreviated to $A$, $B$, $C$, and "$t$`` is used as the abbreviation of time bounds due to the space limitation.}
 \label{result:traj}
\end{figure*}
In this section, we investigate the performance of the proposed encoder-decoder structured NN controller through a numerical experiment. 
In particular, we compare the control performance among the sequential, tree-structured, and graph-structured encoders as explained in Section~\ref{sec:proposed} using a numerical experiment of a path planning problem in the 2D space. 
All of the experiments are conducted in Python running on a Windows~10 with a 2.80 GHz Core i7 CPU and 32 GB of RAM. The NNs are implemented using PyTorch \cite{pytorch}, which is an open-source library for machine learning and, in particular, we use PyTorch geometric \cite{geometric} for the implementation of GNNs. 
We also used the STLCG toolbox \cite{stlcg} for the computation of STL robustness.

We consider a unicycle car-like robot with the following dynamics:
\begin{align}\label{dyn}
    [\dot{q}_x\ \dot{q}_y\ \dot{\theta} ]^\top = [v\cos \theta\ v\sin \theta \ \omega]^\top,
\end{align}
where $[q_x, q_y]$ represents the 2-D position of the vehicle, $\theta$ is the heading angle, and $v$, $\omega$ are the velocity and angular velocity of the vehicle, respectively. 
The system state $x$ and control input $u$ are defined by $x = [q_x,q_y,\theta]^\top$ and $u = [v,\omega]^\top$ respectively.
We discretize the continuous-time dynamics (\ref{dyn}) by a zero-order hold to obtain the discrete-time dynamics. Then, in each discrete time step $k$, we impose the constraint for the velocity of the vehicle by $|v_k|<1.5$ and set the horizon length of the control problem to $T=16$.
For the decoder, we use 2-layered LSTM with 32-dimensional hidden states for the entire experiment same as the works, 
and for the encoder we compare the performance among the following three structures: (I) 1-layered LSTM with 128-dimensional hidden states; 
(II) GNN with 128-dimensional node embeddings, hyperbolic tangent activation for the aggregation step, the number of aggregation step $K=3$, and max-pooling graph embedding; (III) Tree LSTM with 128-dimensional hidden states. 
The dimension for the embedding layer in models (I) and (III) is set to 32 (see Fig \ref{fig:proposed}).
Since the dimension of the hidden states of the encoder and decoder models are different, we input the vector generated by the encoder to the decoder with the state of the system in each time step instead of defining the initial hidden state of the decoder as the vector received from the encoder.
Moreover, we test the performance with and without the attention mechanisms for all cases.
The parameters are updated by Adam optimizer \cite{adam} and the learning rate is set to 0.0003. When we input the STL specification to the encoder, we use vectors defined in Table \ref{table:vector}. 
\begin{table}[tb]
 \caption{Vectors assigned to all the elements in STL specification}
 \label{table:vector}
 \centering
  \scalebox{1}{\begin{tabular}{ll}
   \hline
   Marks & Vector representation \\
   \hline \hline
   $\lnot$ (negation)&$[1,0,0,0,0,0,0,0,0,0,0,0,0,0]^\top$\\
   $\land$ (conjunction)& $[0,1,0,0,0,0,0,0,0,0,0,0,0,0]^\top$\\
   $\lor$ (disjunction) &  $[0,0,1,0,0,0,0,0,0,0,0,0,0,0]^\top$ \\
   $\bm{F}$ (eventually) & $[0,0,0,1,0,0,0,0,0,0,0,0,0,0]^\top$ \\
   $\bm{G}$ (always) & $[0,0,0,0,1,0,0,0,0,0,0,0,0,0]^\top$ \\
   $\bm{U}$ (until) & $[0,0,0,0,0,1,0,0,0,0,0,0,0,0]^\top$ \\
   ( (left bracket)& $[0,0,0,0,0,0,1,0,0,0,0,0,0,0]^\top$\\
   ) (right bracket) & $[0,0,0,0,0,0,0,1,0,0,0,0,0,0]^\top$\\
   $[\tau,\tau']$ (time interval) & $[0,0,0,0,0,0,0,0,\tau,\tau',0,0,0,0]^\top$\\  
   $x_t\in[a,b]\times [c,d]$  &$[0,0,0,0,0,0,0,0,0,0,a,b,c,d]^\top$\\

   \hline
  \end{tabular}}
\end{table}

The specifications that we consider in this experiment are
summarized in Table \ref{table:spe}. 
\begin{table}[tb]
 \caption{Template specifications}
 \label{table:spe}
 \centering
  \scalebox{0.9}{\begin{tabular}{c|l}
   \hline
   & Specification templates \\
   \hline \hline
   (T1)& $((\{\bm{G}/\bm{F}\}_{[\tau_1,\tau'_1]}(x_t\in \mathrm{A}))\{\lor / \land\} (\{\bm{G}/\bm{F}\}_{[\tau_2,\tau'_2]}(x_t\in\mathrm{B})))$\\
&$\{\lor / \land\} (\{\bm{G}/\bm{F}\}_{[\tau_3,\tau'_3]}(x_t\in\mathrm{C}))$\\
(T2)& $(\{\bm{G}/\bm{F}\}_{[\tau_1,\tau'_1]}(x_t\in \mathrm{A}))\{\lor / \land\}  ((\{\bm{G}/\bm{F}\}_{[\tau_2,\tau'_2]}(x_t\in\mathrm{B})$\\
&$\{\lor / \land\} \{\bm{G}/\bm{F}\}_{[\tau_3,\tau'_3]}(x_t\in\mathrm{C})))$\\
(T3)& $((\bm{F}_{[0,T-\tau_4]}\bm{G}_{[0,\tau_4]}(x_t\in \mathrm{A}))\{\lor / \land\} (\bm{F}_{[0,T-\tau_5]}\bm{G}_{[0,\tau_5]}(x_t\in\mathrm{B})))$ \\
&$\{\lor / \land\} (\bm{F}_{[0,T-\tau_6]}\bm{G}_{[0,\tau_6]}(x_t\in\mathrm{C}))$\\
(T4)& $(\bm{F}_{[0,T-\tau_4]}\bm{G}_{[0,\tau_4]}(x_t\in \mathrm{A}))\{\lor / \land\} ((\bm{F}_{[0,T-\tau_5]}\bm{G}_{[0,\tau_5]}(x_t\in\mathrm{B}))$ \\
&$\{\lor / \land\} (\bm{F}_{[0,T-\tau_6]}\bm{G}_{[0,\tau_6]}(x_t\in\mathrm{C})))$\\
   
   \hline
  \end{tabular}}
\end{table}
Here, $\tau_i$ and $\tau'_i$ are the randomly chosen integers with $\tau_i<\tau'_i$ and $\tau'_i \leq T=16$ (for the case $i=1,2,3$) and $\tau_i\leq 8$ (for the case $i=4,5,6$). $\mathrm{A}$, $\mathrm{B}$, and $\mathrm{C}$ are $1.5\times 1.5$ square regions randomly sampled from the region [0,5]$\times$[0,5] without overlapping (see Figure \ref{result:traj}). The slash ``/"  means that we consider both specifications that are separately defined by the right- and left-hand sides of components (i.e., $\{\bm{F}/\bm{G}\}_{[\tau,\tau']} (x_t \in \mathrm{A})$ means that we consider both specifications $\bm{F}_{[\tau,\tau']}(x_t \in \mathrm{A})$ and $\bm{G}_{[\tau,\tau']}(x_t\in\mathrm{A}$)).
For simplicity, if the former $\{\lor/\land\}$ in (T1)-(T4) is defined by $\land$, the letter $\{\lor/\land\}$ is automatically defined by $\lor$. The total number of variations of the specification structures in Table \ref{table:spe} arising from the combinations of the operators and the positions of the bracket pairs is 50. 
Moreover, since regions A, B, and C are sampled from continuous space, the total number of the target specifications in $\Phi$ in this example is infinite.
The initial state is fixed and set to $x_0=[0,0,0]^\top$.

For instance, the specifications $\bm{F}_{[\tau,\tau']}(x_t \in \mathrm{A})$, $\bm{G}_{[\tau,\tau']}(x_t\in\mathrm{A})$ and $\bm{F}_{[0,T-\tau]}\bm{G}_{[0,\tau]}(x_t\in \mathrm{A})$ mean, ``reach $\mathrm{A}$ within the time interval $[\tau_1,\tau'_1]$", ``stay $\mathrm{A}$ within the time interval $[\tau_1,\tau'_1]$", and ``stay $\mathrm{A}$ $\tau$ time steps within the time interval $[0,T]$", respectively. 
In each training iteration, one specification is sampled for every type of template by randomly choosing time bounds and regions. Then, we use them to calculate the loss (\ref{ave robustness}) and update the NN parameters. 
The specifications used for testing the control performance are sampled in the same way as the specifications for the training but only the ones that are confirmed to be satisfiable (i.e., the corresponding control problem is confirmed to be feasible) by the solver are collected.
In this experiment, 200 specifications are collected for each template and used to test the control performance.

The results of the experiment are shown in Fig. \ref{result:robustness1}, \ref{result:robustness2}, \ref{result:traj},  and Table \ref{result:whole}, \ref{result:each}.
Fig. \ref{result:robustness1} and \ref{result:robustness2} show the negative averaged robustness and success rate (the rate of the trajectories that achieve positive robustness) for the test specifications across the training iterations obtained by using the controllers with the sequential, graph-structured and tree-structured encoders (Fig. \ref{result:robustness1} and \ref{result:robustness2} are the results for the without and with attention mechanism, respectively). From Fig \ref{result:robustness1}, we can see that the results of the sequential encoder without attention mechanism are inferior to the other encoder structures. The main reason for this is that the sequential encoder suffers from reading the operator's range of influence (e.g., the sequential model could not distinguish the difference between the specifications $((\bm{F}_{[\tau_1,\tau'_1]}(x_t\in \mathrm{A}))\lor (\bm{G}_{[\tau_2,\tau'_2]}(x_t\in\mathrm{B})))\land (\bm{G}_{[\tau_3,\tau'_3]}(x_t\in\mathrm{C}))$ and $(\bm{F}_{[\tau_1,\tau'_1]}(x_t\in \mathrm{A}))\lor ((\bm{G}_{[\tau_2,\tau'_2]}(x_t\in\mathrm{B}))\land (\bm{G}_{[\tau_3,\tau'_3]}(x_t\in\mathrm{C})))$  well). On the other hand, the tree-structured and graph-structured encoders achieve much better control performance (the performance of the tree-structured encoder is superior to that of the other encoder structures). 
Fig \ref{result:robustness2}, shows that the tree-structured encoder with an attention mechanism achieves faster convergence and relatively higher control performance than the sequential encoder while the graph-structured encoder with an attention mechanism does not work well in this experiment. 
Moreover, we can also see that the performance of the controller with the sequential encoder is quite improved by employing the attention mechanism. 
\begin{table}[tb]
  \caption{Averaged robustness and success rate finally attained by each controller.}
  \label{result:whole}
  \centering
  \begin{tabular}{lcc}
    \hline
     \multicolumn{3}{c}{without attention mechanism}\\
    \hline
    
     Encoder type & Averaged robustness & Success rate\\
     
     \hline
     (I) & 0.23 & 0.62\\
     (II) & 0.46 & 0.89\\
     (III) & 0.50 & 0.92\\
     \hline
     \hline
     \multicolumn{3}{c}{with attention mechanism}\\
     \hline
    
     Encoder type & Averaged robustness & Success rate\\
     
     \hline
     (I) & 0.51 & 0.91\\
     (II) & 0.27 & 0.70\\
     (III) & 0.52 & 0.93\\
     
    \hline
  \end{tabular}
\end{table}

In Table \ref{result:whole}, we summarize the averaged robustness and success rate finally attained by each encoder type. The tree-structured encoder with the attention mechanism shows the highest performance and we have confirmed that over 93 \% of the test specifications are satisfied (similar performances are achieved by the sequential encoder with the attention mechanism and tree-structured encoder without the attention mechanism).
In Table \ref{result:each}, we also summarize the negative averaged robustness and success rate for each template specification. Note that the specifications in Table \ref{result:each} are indicated by omitting the time bounds due to the space limitation and the results for the specifications that have the same meanings (e.g., $(\bm{F}(x_t\in \mathrm{A}))\land ((\bm{F}(x_t\in\mathrm{B}))\lor (\bm{G}(x_t\in\mathrm{C})))$, $((\bm{G}(x_t\in\mathrm{A}))\lor(\bm{F}(x_t\in \mathrm{B})))\land (\bm{F}(x_t\in\mathrm{C}))$, $(\bm{F}(x_t\in \mathrm{A}))\land ((\bm{G}(x_t\in\mathrm{B}))\lor (\bm{F}(x_t\in\mathrm{C})))$, and $((\bm{F}(x_t\in\mathrm{A}))\lor(\bm{G}(x_t\in \mathrm{B})))\land (\bm{F}(x_t\in\mathrm{C}))$) are collectively displayed.

Fig. \ref{result:traj} shows a few examples of the vehicle trajectories for some test specifications generated by applying the trained control policy to the system (above) and the visualization of the attention weights for the controller with sequential encoder (below). From this figure, we can see that the vehicle is flexibly controlled to satisfy the given specifications by appropriately distinguishing the operators meaning and considering time constraints. 
Moreover, we can see the relatively intuitive attention weights from the figure that high attention weights are assigned to the part representing regions to be visited. 
\begin{table}[tb]
 \caption{Template specifications}
 \label{table:spe2}
 \centering
  \scalebox{0.9}{\begin{tabular}{c|l}
   \hline
   & Specification templates 2 \\
   \hline \hline
   (T'1)& $\{\bm{G}/\bm{F}\}_{[\tau_1,\tau'_1]}(x_t\in \mathrm{A})$\\

(T'2)& $(\{\bm{G}/\bm{F}\}_{[\tau_1,\tau'_1]}(x_t\in \mathrm{A}))\{\land / \lor \}(\{\bm{G}/\bm{F}\}_{[\tau_2,\tau'_2]}(x_t\in \mathrm{B}))$\\

(T'3)& $((\bm{F}_{[0,T-\tau_4]}\bm{G}_{[0,\tau_4]}(x_t\in \mathrm{A}))\{\lor / \land\} (\{\bm{G}/\bm{F}\}_{[\tau_1,\tau'_1]}(x_t\in \mathrm{B})))$\\
&$\{\land / \lor \}(\{\bm{G}/\bm{F}\}_{[\tau_2,\tau'_2]}(x_t\in \mathrm{C}))$ \\

   \hline
  \end{tabular}}
\end{table}
\begin{table}[tb]
  \caption{Averaged robustness and success rate for the specifications (T'1)-(T'3).}
  \label{result:additional}
  \centering
  \begin{tabular}{lcc}
    \hline
     \multicolumn{3}{c}{(I) with attention mechanism}\\
    \hline
    
     Specification type & Averaged robustness & Success rate\\
     
     \hline
     (T'1) & 0.59 & 1.00\\
     (T'2) & 0.42 & 0.85\\
     (T'3) & 0.37 & 0.81\\
     \hline
     \hline
     \multicolumn{3}{c}{(III) with attention mechanism}\\
     \hline
    
     Specification type & Averaged robustness & Success rate\\
     
     \hline
     (T'1) & -0.35 & 0.32\\
     (T'2) & -0.18 & 0.56\\
     (T'3) & 0.44 & 0.89\\
     
    \hline
  \end{tabular}
\end{table}

Lastly, we test the performance of the controllers trained with the templates in Table \ref{table:spe} for the test specifications generated from the different templates (T'1)-(T'3) in Table \ref{table:spe2}. The numbers of variations of templates in (T'1)-(T'3) are 2, 6, and 12, respectively.
A set of test specifications is constructed by collecting 200 specifications for each template same as the testing for (T1)-(T4). 
The results are shown in Table \ref{result:additional}. 
For the templates in (T'1) and (T'2), the controller with the sequential encoder meets the specifications with a high rate without additional adaptation steps while the tree-structured encoder does not work well in this example. This result may be from the fact that the structures of the parse trees for the training specifications are quite different from that of the testing specifications.
On the other hand, For the template (T'3), the result of the tree-structured encoder is better than that of the sequential encoder.
From these results, we can see that the proposed method potentially can deal with the specifications sampled from the different distribution (in this case, different template structures) from the one considered in the training while which encoder structure achieves better performance would be depending on the specifications considered in the training and testing. 

In summary, we have observed the followings from all of the above results. First, the proposed method enables the construction of NN controllers that can generate control inputs satisfying a wide range of STL specifications with various time constraints, formula structures, and changing predicates. Second, the encoder NN architecture is an important factor for encoding STL. Specifically, we have mainly seen the following characteristics: (i) Compared to the sequential encoder, the tree-structured encoder enables better control performance and faster convergence. This result suggests that the tree-structured encoder can better capture the logical structure of STL than the sequential one. (ii) The attention mechanism is effective for both the sequential and tree-structured encoders but not for the graph-structured encoder. Especially, for the sequential encoder, the control performance finally attained is much superior to that of the sequential encoder without the attention mechanism. (iii) We have found that the performance of the tree-structured encoder is basically superior to that of the graph-structured encoder. One reason for this may be that the bottom-up process of the tree-structured NN is suited for our problem compared to the synchronized process of the GNNs. Lastly, we have confirmed the potential of the proposed method to deal with the specifications sampled from a different distribution from the one considered in the training.
\section{Conclusion and Future Direction}
In this paper, we proposed a way to generalize NN-based STL control synthesis using the encoder-decoder NN architectures and concrete training procedures of them. The encoder was constructed by a sequential, graph-structured, or tree-structured NNs to read and encode STL specifications given by the user while the decoder that outputs control inputs based on the vector obtained from the encoder and state of the system was constructed by a sequential NN that can deal with the history-dependent nature of the STL specification satisfaction. The attention mechanism is employed to further improve the control performance.
All of the parameters within the proposed NN controller were trained in an end-to-end manner to maximize the expected robustness for the initial states and STL specifications sampled from the given distribution.
The result from the case study presented in Section \ref{case study} showed the efficacy of the proposed method. 

As a future direction of this work, we will explore an advanced NN architecture that can more flexibly extract the crucial information from the STL specification and encode it to latent representation. Specifically, although, in this work, the time constraints are simply encoded into the vector in Table \ref{table:vector} and fed to the encoder, a more elegant way to incorporate such information into NN would be an interesting study direction.
Moreover, as commonly discussed in the STL control synthesis literature, local optima that do not satisfy the given specification are likely to be obtained especially when we consider the control problem with complex specifications including nested operators because of the non-convexity of the smooth robustness function. This problem potentially narrows the applicability of our method.
One possible way to avoid this problem is to use expert demonstrations to guide the training as discussed in \cite{semi}. However, since the problem considered in this paper require us to train the controller for many types of STL specifications, a very large amount of expert trajectories will be needed if we simply use such a method. Thus, finding a more realistic solution to the local optima problem will be another future direction of this work.
\section*{Acknowledgement}
This work is supported by JST CREST JPMJCR201, Japan and by JSPS KAKENHI Grant 21K14184.

\appendix
Here, we summarize the detailed architecture of the tree-structured encoder discussed in Section \ref{encoder}. To this end, we first explain the detailed update equations of the original LSTM proposed in \cite{LSTM} and then proceed to the explanation of the Tree LSTM \cite{tree} based encoder model.
\subsection{LSTM}\label{ape:lstm}
The LSTM is the specific type of sequential NNs, which employs \textit{memory cell} $c_t$ that can preserve long-term information regarding input sequence \cite{LSTM}. 
The concrete transition equations of the LSTM are as follows:
\begin{subequations}\label{lstm update}
\begin{align}
    i_t &= \tilde{\sigma} \left(W^{(i)}s_t+U^{(i)}h_{t-1}+b^{(i)}\right),\\
    f_t &= \tilde{\sigma} \left(W^{(f)}s_t+U^{(f)}h_{t-1}+b^{(f)}\right),\\
    o_t &= \tilde{\sigma} \left(W^{(o)}s_t+U^{(o)}h_{t-1}+b^{(o)}\right),\\
    u_t &= \mathrm{tanh} \left(W^{(u)}s_t+U^{(u)}h_{t-1}+b^{(u)}\right),\\
    c_t &= i_t\odot u_t + f_t\odot c_{t-1},\\
    h_t &= o_t \odot \mathrm{tanh}(c_t),
\end{align}
\end{subequations}
where $s_t$ is the input at time $t$ (in our case, the vector that represents each component within the given STL formula), $\tilde{\sigma}$ denotes logistic sigmoid activation function, and $\odot$ denotes element-wise multiplication. $i_t$, $f_t$, $o_t$, and $u_t$ are so-called an \textit{input gate}, \textit{forget gate}, \textit{output gate}, and \textit{hidden state}, respectively. $W^{(i)}$, $W^{(o)}$, $W^{(f)}$, $W^{(u)}$, $b^{(i)}$, $b^{(o)}$, $b^{(f)}$, $b^{(u)}$ are the parameters to be trained.
We note here that the concatenation of the hidden state $h$ and memory cell $c$ in above is regarded as the hidden state in Section \ref{encoder}.
\subsection{Tree LSTM}\label{ape:treelstm}
The tree LSTM proposed in \cite{tree} is a type of LSTM that allows tree-structured information propagation.
In this study, we employ the model structure based on the Child-Sum Tree LSTM in \cite{tree} (minor modification is made in (\ref{update h}) to deal with the irreversible nature of the \textit{until} operator). The concrete transition questions are as the followings:
\begin{subequations}

\begin{align}
    \hat{h}_j &= \sum_{k\in C(j)}\sum_{r\in R} W_r h_k\label{update h}\\
    i_j &= \tilde{\sigma} \left(W^{(i)}s_j+U^{(i)}\hat{h}_{j}+b^{(i)}\right)\\
    f_{jk} &= \tilde{\sigma} \left(W^{(f)}s_j+U^{(f)}h_{k}+b^{(f)}\right)\\
    o_j &= \tilde{\sigma} \left(W^{(o)}s_j+U^{(o)}\hat{h}_{j}+b^{(o)}\right)\\
    u_j &= \mathrm{tanh} \left(W^{(u)}s_j+U^{(u)}\hat{h}_{j}+b^{(u)}\right)\\
    c_j &= i_j\odot u_j + \sum_{k\in C(j)}f_{jk}\odot c_{k}\label{update c}\\
    h_j &= o_j \odot \mathrm{tanh}(c_j)
\end{align}
\end{subequations}
where $C(j)$ represents the set of children nodes of node $j$ and $R=\{0,1,2\}$ represents the set of relations ($1$ is assigned to the incoming edge directed to the right-hand side of the \textit{until} operator, $1$ is assigned to the incoming edge directed to the left-hand side of the \textit{until} operator, and $0$ is assigned to the other relations). The difference from the ordinary LSTM update equations (\ref{lstm update}) is that the hidden state and the memory cell are updated based on that of the children nodes as (\ref{update h}) and (\ref{update c}). 
\newpage

\begin{landscape}
\begin{table}[h]
\caption{Negative averaged robustness and success rate for all the template specifications finally attained by each model structure.}
\label{result:each}
\begin{tabular}{l|r|r|r|r|r|r|r|r|r|r|r|r|}
\hline
{} & \multicolumn{2}{|c|}{LSTM} &  \multicolumn{2}{|c|}{GNN} &  \multicolumn{2}{|c|}{Tree LSTM}& \multicolumn{2}{|c|}{Attention LSTM} &  \multicolumn{2}{|c|}{Attention GNN} &  \multicolumn{2}{|c|}{Attention Tree LSTM}\\
\hline
Specification type &  Robustness &  Success &  Robustness &  Success &  Robustness &  Success &  Robustness &  Success&  Robustness &  Success &  Robustness &  Success \\
\hline \hline
1. \scalebox{0.7}{$(\bm{F}(x_t\in \mathrm{A}))\land ((\bm{F}(x_t\in\mathrm{B}))\lor (\bm{F}(x_t\in\mathrm{C})))$ } &             0.327976 &      0.040000 &            -0.301471 &      0.840000 &            -0.423321 &      0.960000 &            -0.541322 &      0.975000 &            -0.227308 &       0.86000 &            -0.474158 &      0.940000 \\
2. \scalebox{0.7}{$((\bm{F}(x_t\in \mathrm{A}))\land (\bm{F}(x_t\in\mathrm{B})))\lor (\bm{F}(x_t\in\mathrm{C}))$ }  &            -0.696736 &      1.000000 &            -0.626746 &      1.000000 &            -0.704850 &      1.000000 &            -0.713727 &      0.995000 &            -0.675400 &       1.00000 &            -0.694092 &      1.000000 \\
3. \scalebox{0.7}{$(\bm{F}(x_t\in \mathrm{A}))\lor (\bm{F}(x_t\in\mathrm{B}))\lor (\bm{F}(x_t\in\mathrm{C}))$ }  &            -0.690055 &      0.995098 &            -0.633360 &      1.000000 &            -0.702956 &      1.000000 &            -0.710401 &      1.000000 &            -0.669008 &       1.00000 &            -0.689601 &      1.000000 \\
4. \scalebox{0.7}{$(\bm{F}(x_t\in \mathrm{A}))\land ((\bm{F}(x_t\in\mathrm{B}))\lor (\bm{G}(x_t\in\mathrm{C})))$ }  &             0.381445 &      0.030000 &            -0.284011 &      0.840000 &            -0.417159 &      0.920000 &            -0.448311 &      0.925000 &            -0.167249 &       0.80000 &            -0.465502 &      0.920000 \\
5. \scalebox{0.7}{$((\bm{F}(x_t\in \mathrm{A}))\land (\bm{F}(x_t\in\mathrm{B})))\lor (\bm{G}(x_t\in\mathrm{C}))$ }  &            -0.103956 &      0.450000 &            -0.254708 &      0.860000 &            -0.364633 &      0.920000 &            -0.347309 &      0.815000 &            -0.144392 &       0.76000 &            -0.249271 &      0.820000 \\
6. \scalebox{0.7}{$((\bm{F}(x_t\in \mathrm{A}))\lor (\bm{F}(x_t\in\mathrm{B})))\land (\bm{G}(x_t\in\mathrm{C}))$ }   &             0.656914 &      0.065000 &             0.100892 &      0.500000 &            -0.155257 &      0.640000 &            -0.103916 &      0.670000 &             0.591498 &       0.08000 &            -0.213135 &      0.760000 \\
7. \scalebox{0.7}{$(\bm{F}(x_t\in \mathrm{A}))\lor ((\bm{F}(x_t\in\mathrm{B}))\land (\bm{G}(x_t\in\mathrm{C})))$ }  &            -0.701816 &      1.000000 &            -0.617034 &      1.000000 &            -0.700855 &      1.000000 &            -0.721272 &      1.000000 &            -0.680011 &       1.00000 &            -0.693411 &      1.000000 \\
8. \scalebox{0.7}{$(\bm{F}(x_t\in \mathrm{A}))\lor (\bm{F}(x_t\in\mathrm{B}))\lor (\bm{G}(x_t\in\mathrm{C}))$ } &            -0.692943 &      1.000000 &            -0.583235 &      1.000000 &            -0.652931 &      1.000000 &            -0.706845 &      1.000000 &            -0.652286 &       1.00000 &            -0.654281 &      1.000000 \\
9. \scalebox{0.7}{$(\bm{F}(x_t\in \mathrm{A}))\land ((\bm{G}(x_t\in\mathrm{B}))\lor (\bm{G}(x_t\in\mathrm{C})))$ }  &             0.683614 &      0.000000 &             0.319070 &      0.200000 &            -0.162956 &      0.600000 &            -0.166623 &      0.710000 &             0.587816 &       0.02000 &            -0.015224 &      0.560000 \\
10. \scalebox{0.7}{$((\bm{F}(x_t\in \mathrm{A}))\land (\bm{G}(x_t\in\mathrm{B})))\lor (\bm{G}(x_t\in\mathrm{C}))$ }  &            -0.402167 &      0.780000 &            -0.491718 &      0.860000 &            -0.454669 &      0.820000 &            -0.375936 &      0.785000 &            -0.474956 &       0.84000 &            -0.517188 &      0.860000 \\
11. \scalebox{0.7}{$((\bm{F}(x_t\in \mathrm{A}))\lor (\bm{G}(x_t\in\mathrm{B})))\land (\bm{G}(x_t\in\mathrm{C}))$ } &             0.606332 &      0.020000 &             0.050534 &      0.540000 &            -0.211490 &      0.780000 &            -0.180167 &      0.705000 &             0.569004 &       0.04000 &            -0.221290 &      0.800000 \\
12. \scalebox{0.7}{$(\bm{F}(x_t\in \mathrm{A}))\lor ((\bm{G}(x_t\in\mathrm{B}))\land (\bm{G}(x_t\in\mathrm{C})))$ }  &            -0.705601 &      1.000000 &            -0.628041 &      1.000000 &            -0.701515 &      1.000000 &            -0.717200 &      1.000000 &            -0.683054 &       1.00000 &            -0.694087 &      1.000000 \\
13. \scalebox{0.7}{$(\bm{F}(x_t\in \mathrm{A}))\lor (\bm{G}(x_t\in\mathrm{B}))\lor (\bm{G}(x_t\in\mathrm{C}))$} &            -0.679380 &      1.000000 &            -0.595460 &      1.000000 &            -0.635411 &      1.000000 &            -0.700237 &      1.000000 &            -0.634373 &       1.00000 &            -0.661482 &      1.000000 \\
14. \scalebox{0.7}{$(\bm{G}(x_t\in \mathrm{A}))\land ((\bm{G}(x_t\in\mathrm{B}))\lor (\bm{G}(x_t\in\mathrm{C})))$ } &             0.397373 &      0.000000 &            -0.102460 &      0.620000 &            -0.277550 &      0.760000 &            -0.173673 &      0.725000 &             0.283947 &       0.20000 &            -0.278279 &      0.800000 \\
15. \scalebox{0.7}{$((\bm{G}(x_t\in \mathrm{A}))\land (\bm{G}(x_t\in\mathrm{B})))\lor (\bm{G}(x_t\in\mathrm{C}))$ } &            -0.499990 &      0.855000 &            -0.487519 &      0.940000 &            -0.564431 &      0.920000 &            -0.478985 &      0.900000 &            -0.472312 &       0.88000 &            -0.527234 &      0.940000 \\
16. \scalebox{0.7}{$(\bm{G}(x_t\in \mathrm{A}))\lor (\bm{G}(x_t\in\mathrm{B}))\lor (\bm{G}(x_t\in\mathrm{C}))$ } &            -0.603754 &      0.975490 &            -0.425490 &      0.882353 &            -0.400755 &      0.931373 &            -0.606620 &      0.955882 &            -0.471917 &       0.95098 &            -0.474357 &      0.941176 \\
17. \scalebox{0.7}{$(\bm{FG}(x_t\in \mathrm{A}))\land ((\bm{FG}(x_t\in\mathrm{B}))\lor (\bm{FG}(x_t\in\mathrm{C})))$ } &             0.388273 &      0.000000 &            -0.474170 &      1.000000 &            -0.483941 &      0.960000 &            -0.553777 &      0.990000 &             0.043371 &       0.46000 &            -0.505987 &      0.980000 \\
18. \scalebox{0.7}{$((\bm{FG}(x_t\in \mathrm{A}))\land (\bm{FG}(x_t\in\mathrm{B})))\lor (\bm{FG}(x_t\in\mathrm{C}))$ } &            -0.678117 &      1.000000 &            -0.623349 &      1.000000 &            -0.703507 &      1.000000 &            -0.695220 &      1.000000 &            -0.653996 &       1.00000 &            -0.689870 &      1.000000 \\
19. \scalebox{0.7}{$(\bm{FG}(x_t\in \mathrm{A}))\lor (\bm{FG}(x_t\in\mathrm{B}))\lor (\bm{FG}(x_t\in\mathrm{C}))$ } &            -0.631755 &      0.985294 &            -0.621294 &      1.000000 &            -0.696392 &      1.000000 &            -0.606398 &      0.975490 &            -0.641196 &       1.00000 &            -0.676629 &      1.000000 \\
\hline
\end{tabular}
\end{table}
\end{landscape}

\end{document}